%Paper: 9204005
%From: Jonathan Katz <katz@wuphys.wustl.edu>
%Date: Thu, 30 Apr 1992 14:18:05 -0500

\let\rawfootnote=\footnote
\def\footnote#1#2{{\baselineskip=\normalbaselineskip
    \parindent=0pt\parskip=0pt\rm
    \rawfootnote{#1}{#2\hfill\vrule height 0pt depth 6pt width 0pt}}}
\magnification=\magstep1
\baselineskip=20pt
\centerline{Radiation Transfer in Gamma-Ray Bursts}
\bigskip
\centerline{B. J. Carrigan\footnote*{Department of Physics and McDonnell
Center for the Space Sciences, Washington U., St.~Louis, Mo.
63130}\footnote\dag{Present Address: Department of Physics, Millikin U.,
Decatur, Ill. 62522} and J. I. Katz$^\ast$}
\bigskip
\centerline{Abstract}
\bigskip
We have calculated gamma-ray radiative transport in regions of high energy
density, such as gamma-ray burst source regions, using a discrete ordinate,
discrete energy group method.  The calculations include
two-photon pair production and annihilation, as well as three-photon pair
annihilation.  The radiation field itself acts as an absorbing medium, and
the optical depth depends on its intensity, so the problem is intrinsically
nonlinear.  Spherical divergence produces effective collimation of the flux.
At high optical depth the high energy ($E > 1$ MeV) portion of the emergent
spectrum assumes a nearly universal form.  An approximate limit is derived
for the high energy flux from a gamma-ray burst source region of given size,
and the implications of this limit for the distance to the March 5, 1979
event are briefly discussed.  We discuss more generally the problem of very
luminous bursts, and implications of Galactic halo distances for flare
models.
\par
\vfil
\noindent
Subject headings: Gamma Rays: Bursts, Radiative Transfer
\par
\eject
\centerline{I. INTRODUCTION}
\bigskip
Gamma-ray bursts pose an unusual and interesting problem in radiation
transport.  In regions of low matter density, but high gamma-ray radiation
density, the chief source of opacity at gamma-ray ($> {1 \over 2}$ MeV)
energy is pair production by interaction with other gamma-rays.  The opacity
is proportional to the radiation energy density, making the radiation
transport equation intrinsically nonlinear.  Because of the kinematics of
two-photon pair production, the opacity is also a sensitive function of the
angular and spectral distribution of the radiation field.

The purpose of this paper is to describe the results of gamma-ray radiation
transport calculations under the influence of pair-production opacity,
making the approximation that other sources of opacity (such as Compton
scattering) are negligible.  A preliminary account of our work, restricted
to the case of slab-symmetry, was reported by Carrigan and Katz (1984).
Because of the importance of the photon angular distribution to the opacity,
spherical divergence is essential to a realistic calculation, and it is
used in the calculations reported here.

These radiation transport calculations are also of interest because arguments
based on pair-production opacities have been used to constrain the distances of
gamma-ray burst sources.  Direct estimates of their distances are
controversial, and range from $< 100$ pc to cosmological values.  There are no
universally accepted identifications of burst sources with quiescent
counterparts whose distances might be determined by independent astronomical
arguments.  Statistical arguments, using empirical $\log N / \log S$
relations, have a long history of controversy, because of questions of
completeness of sampling at lower flux levels.  This controversy may be
resolved, at least for the more common classes of burst sources, by data
from the BATSE on GRO, but some of the most interesting sources, such as the
soft gamma repeaters (SGR), are few enough to preclude statistical analysis.
The SGR also have a large range of intrinsic brightness, complicating $\log
N / \log S$ analysis, but excluding models (such as thermonuclear) which
predict a ``standard candle'' of luminosity.

Perhaps the most interesting problem is presented by GBS 0526-66 (the March 5,
1979, source; Cline 1980), whose position on the sky coincides with the
supernova remnant N 49.  However, N 49 lies in the Large
Magellanic Cloud, so the inferred luminosity of bursts
from this source is huge, and many scientists have been
reluctant to accept the identification of N 49 with GBS
0526-66.  Faced with this lack of information, attempts
have been made to obtain an upper bound on burster
distances from the observed burst spectra.

     The argument (Schmidt 1978, Cavallo and Rees 1978, Katz 1982, Epstein
1985) proceeds as follows: two photons which
are above the threshold for electron-positron pair production
$$E_1 E_2 (1-\cos\theta_{12}) \geq 2 (m_e c^2)^2 \eqno(1)$$
may produce a pair, where $\theta_{12}$ is the angle between the directions
of the two gamma-rays.  The cross section for pair production
reaches a maximum at a finite CM photon energy (roughly 700 KeV); if the
photon spectrum is not sharply peaked high-energy photons will, therefore,
form pairs predominantly with low-energy photons.
Because any reasonable source spectrum will contain many
more low- or moderate-energy photons than high-energy
photons, the emergent spectrum will differ most markedly
from the source spectrum at high photon energies ($E > 1$ MeV), at which
it will be heavily depleted.   The electron-positron pairs eventually
annihilate to produce two (infrequently three) photons, but usually
not one high- and one low-energy photon.  The result is
that high-energy photons are preferentially removed from
the spectrum.  The observation of a measurable amount of
flux with $E > 1$ MeV is thus not expected unless the optical
depth to pair production is unity or less.  However, the
optical depth is given approximately by
$$\tau \sim n_\gamma r_e^2 R, \eqno(2)$$
where $n_\gamma$ is the number density, at the source, of photons
with energy above threshold for pair production off the
observed high-energy photons, $r_e$ is the classical electron radius
$e^2/(m_e c^2)$ (the cross sections for pair production
and annihilation, as well as for Compton scattering, are all $\sim r_e^2$
at these semi-relativistic energies), and $R$ is the radius of the
source.  The condition of low ($\tau < 1$) optical depth then yields
$$n_\gamma < r_e^{-2} R^{-1} \eqno(3)$$
or, for the flux at the source,
$$F_0 \sim n_\gamma c < c r_e^{-2} R^{-1}. \eqno(4)$$
The flux at the solar system, a distance $d$ from the source, is
$$F(d) = F_0 (R/d)^2 < c R d^{-2} r_e^{-2}, \eqno(5)$$
and hence
$$d < [c R F(d)^{-1} r_e^{-2}]^{1/2}, \eqno(6)$$
or
$$d < 200 {\rm pc} (R/10 {\rm km})^{1/2} [F(d)/1 {\rm cm}^{-2} {\rm
s}^{-1}]^{-1/2}. \eqno(7)$$

For a typical burst, the peak flux implies an upper bound
to the source distance of a few hundred to a few thousand
parsecs.  For the intense March 5, 1979 event, however, the
distance limit inferred in this way from the peak flux is $\sim 30$ pc;
in addition, the presence of an emission
feature in the initial 4-second spectrum of this burst implies
a tighter bound of $\sim 7$ pc (based on a Comptonization argument);
even tighter bounds of 4.3 pc and
$< 1$ pc (Helfand and Long 1979, Katz 1982) have been proposed.  The
reality of apparent emission features in burst spectra
has been questioned, however (Fenimore, {\it et al.} 1982; Nolan, {\it et
al.} 1984; Teegarden 1984; Liang and Petrosian 1986)
so the tightest proposed distance bounds ($< 10$ pc)
may be discounted.  The looser bounds, based
simply on the existence of a high-energy continuum, will
be the only distance bound considered here.  In view of
the tentative localization of the 5 March 1979 source in
the LMC, inconsistent by orders of magnitude with even the loosest of
the bounds mentioned above, it is of interest to investigate
any loopholes in the distance bounds.

     One possible loophole is that the bounds given above are approximate;
a more precise calculation of the cross sections may
provide an optical depth at much larger (or smaller)
fluxes at the source.  A consideration of the exact
pair-production cross section, and a number of reasonable
source spectra, leads to a higher upper bound than the
crude estimate given above.  For the March 5, 1979 source,
the high-energy continuum implies an upper bound of about
150 pc (Epstein 1985).  Though this is a five-fold increase
in the distance bound, it is still far too small to
permit the March 5, 1979 source to be extragalactic.

     A second loophole exists if the source produces a strongly collimated
beam of photons.  In this case, even high energy photons are below
threshold for pair production, because $\theta_{12}$ is small.  A
collimated relativistically-outflowing gas of electrons and positrons
would, on general kinematic grounds, be expected to produce outward
collimated radiation.  Such collimated outflows are inferred in superluminally
expanding active galactic nuclei, and might be expected following any
release of energy density $\gg \rho c^2$, where $\rho$ is the local rest
mass density, but their detailed mechanisms are obscure.

     A final possible loophole begins with a consideration
of all the reactions a system of photons and
charged particles can undergo.  These processes include
Compton scattering, bremsstrahlung, and three- (or more)
photon annihilation.  In addition, the presence of a
magnetic field would allow other processes, such as
synchrotron radiation and one-photon pair production and
annihilation.  Most of these processes are dissipative,
in the sense that the number of particles increase, so
energy is distributed among more particles at the end of
the reaction than at the beginning.  On the other hand,
high-energy photons are also produced.  As a
result, the high-energy flux may be augmented as well as
depleted by interactions, and this may serve to weaken the distance
bounds given above; in other words, a significant high energy flux may
emerge from a region which is optically thick to high energy photons.

There are four ways in which a very intense source of gamma-rays might evade
the bound (6):
\item{1.} Diffusion of radiation through optically thick regions.  However,
because of the proportionality of effective volume opacity to the radiation
energy density, the required increase of energy density with distance into
the opaque region is exponential (as seen in a simple model calculation;
Katz 1982), rather than linear (the usual result for linear radiation
transport).
\item{2.} Repeated pair production and annihilation will, through the
occasional three-body process, lead to a softening of the spectrum.
\item{3.} The presence of a ``window'' in the opacity for collimated photons
suggests that in a region opaque to pair production much of the radiation
may emerge through this window, in analogy to the great contribution of
windows in the material opacity to radiation flow in the usual (Rosseland
mean) approximation.
\item{4.} Outward flow in spherical geometry leads to a purely geometrical
radial collimation of the photon flux (any ray approaches the radial
direction as $r \rightarrow \infty$), increasing the pair-production
threshold.

In order to investigate these processes and to determine how valid
the flux-derived distance bounds are, one must solve the transport
equations in a self-interacting photon system, including the effects of
spherical divergence.  Previous calculations (Carrigan and Katz 1984, 1987;
Guilbert and Stepney 1985) assumed slab-symmetry.  In \S II we discuss the
equations of radiation transport, and in \S III numerical methods for their
solution.  \S IV contains the results of our calculations.  In \S V we
present a summary discussion and conclusions.
\bigskip
\centerline{II. RADIATIVE TRANSPORT}
\bigskip
      The transport equation for a field $\psi$, neglecting scattering, is
$${\partial \psi \over \partial t} + v {\hat \Omega} \cdot \nabla \psi = S -
A \psi, \eqno(8)$$
where $\hat \Omega$ is the unit direction vector, $S$ (the emission
term) is a source of particles (photons, electrons,
or positrons), and $A$ is the absorption coefficient.
In many problems, $S$ and $A$ are independent of
the field, $\psi$.  The transport equation would then be linear,
although $S$ and $A$ often depend on $\psi$ indirectly, through the effects
of $\psi$ on other quantities, such as the temperature in LTE radiative
transport and level occupation densities in non-LTE transport.  The
drift velocity $v$ is very different for photons, which stream freely at the
speed of light between interactions, than for charged particles which may be
confined by a magnetic field.

In the present case, the absorption of
photons arises from photon-photon interactions, while the
absorption of charged particles is due to electron-positron annihilation.
As a result, the absorption
coefficients are proportional to the fields themselves,
so the absorption terms are quadratic in the fields:
$$\eqalign{A_\gamma(E,{\hat \Omega})&=\int \psi_\gamma(E^\prime,
{\hat \Omega}^\prime) Q(E,E^\prime,{\hat \Omega},{\hat \Omega}^\prime)
dE^\prime d{\hat \Omega}^\prime, \cr A_e(E)&=\int \psi_e(E^\prime)
P(E,E^\prime) dE^\prime, \cr} \eqno(9)$$
where $Q$ and $P$ are known functions and $A_\gamma$, $A_e$, $\psi_\gamma$,
and $\psi_e$ implicitly depend on space and time.  In principle, $A_e(E)$,
$\psi_e(E^\prime)$, and $P(E,E^\prime)$ also depend on angle, but this
dependence is not sharp.  Charged particle distributions may be isotropized
by Coulomb scattering, and their pair-production source is not highly
collimated.  Cyclotron radiation appears to anisotropize electron
distributions rapidly, but when emission is rapid so is absorption, with a
cancelling effect, and the net anisotropization rate may not be large.
Because these effects are difficult to calculate quantitatively we
approximate the charged particle distributions as
isotropic, so that $A_e$, $\psi_e$, and $P$ also implicitly include integrals
over their isotropic angular distributions.

The transport equations are unchanged by the following scale transformation:
$$\eqalign{r&\rightarrow \lambda^{-1} r \cr t&\rightarrow \lambda^{-1} t \cr
I \equiv \psi_\gamma &\rightarrow \lambda I \cr n_e \equiv \psi_e
&\rightarrow \lambda n_e \cr S &\rightarrow \lambda^2 S \cr} \eqno(10)$$
A solution of the transport equations for conditions
characteristic of neutron stars may therefore be scaled
to describe transport in other objects, such as active galactic nuclei and
other gamma-ray sources (Matteson 1983, Carrigan and Katz 1987).

The presence of photon-photon pair production makes the transport problem
nonlinear.  One consequence of this is that
we cannot specify the optical depth in advance;
it must be calculated from the solution to
the equations.  As a result, the optical depth
is not a particularly useful parameter in this problem.
Instead, we define $\beta$, which characterizes a family of
solutions (all related by the scale transformation [10]):
$$\beta \equiv {r_e^2 \over R c} \int_{m_e c^2}^\infty dE \int_R^\infty dr
r^2 S(E,r); \eqno(11)$$
$S$ is the source spectrum (the spectrum produced by the primary
radiation process which is assumed to supply the burst's energy), $E$ the
photon energy, and $r$ the radius.  The source region has been assumed to be
the range $R \le r < \infty$, where $R$ might be the neutron star's radius.
A low-energy cut-off in the definition of $\beta$ is required for many
reasonable model source spectra ({\it e.g.}, a power-law) to keep the power
finite; a natural cut-off is 511 keV because lower energy photons are less
effective pair-producers, and we adopted it in (11).
In addition, for a source spectrum which does not
vanish sufficiently rapidly at infinity, a high-energy
cut-off is similarly needed.  In the calculations reported
here, a high-energy cut-off of 2555 keV was used where
such a cut-off was required.  In contrast to the optical
depth,  $\beta$ may be specified at the start of a
transport calculation, and is therefore a useful parameter of the problem.
For small $\beta$, the radiation
field and optical depth are both proportional to $\beta$.

     In spherically symmetric geometry, the transport equation for an
azimuthally symmetric radiant intensity $I(r,t,E,\mu)$ may be written as
$${\partial I \over \partial t} + c\left({\mu \over r^2}{\partial
(r^2 I)\over \partial r} + {1 \over r} {\partial [(1-\mu^2)I] \over \partial
\mu} \right) = G[r,t,E,\mu,I,n_e],\eqno(12a)$$
where $\mu$ is the cosine of the angle between the photon's direction and an
outward radial vector.  The left hand side of (12a) represents the purely
geometrical effects of free-streaming photons.  $G$ includes source and
absorption terms:
$$G[r,t,E,\mu,I,n_e] = n_e^2 \epsilon(T,E) - I(r,t,E,\mu)\int
I(r,t,E^\prime,\mu^\prime) Q(E,E^\prime,\mu,\mu^\prime) dE^\prime
d\mu^\prime + S(E,r). \eqno(12b)$$
The annihilation term is approximated
$$\eqalign{\epsilon(E,T) &= \epsilon_{2\gamma}(E,T) + \epsilon_{3\gamma}(E,T)
\cr &\approx \epsilon_{2\gamma}(E,T) + \alpha \epsilon_{2\gamma}({3 \over 2}
E,T),\cr} \eqno(13)$$
where $\epsilon_{2\gamma}$ is the thermally averaged annihilation spectrum
calculated by Svensson (1983), and the much more complex (and not readily
available) thermally averaged three-photon contribution is roughly
approximated by the scaled two-photon spectrum, as indicated, with $\alpha
\approx 1/137$ being the usual fine structure constant.  Because three-photon
annihilation is small and important only when each photon undergoes many ($>
1/\alpha$) pair production and annihilation cycles this rough approximation
is adequate; its significance is that it increases the number of photons and
particles in the system, degrading their mean energy, and its quantitative
value and form are comparatively unimportant.  Similarly, neglecting Compton
scattering affects lepton-photon energy exchange, both directly by Compton
recoil and indirectly by increasing the effective optical depth and thereby
increasing the effects of other physical processes.

Charged particles are generated only by
pair production.  The electrons and positrons were
assumed to form identical (isotropic) relativistic Maxwell-Boltzmann
distributions; therefore, a number density and temperature
(both functions of $r$ and $t$) suffice to characterize
the charged-particle subsystem.  Isotropy is an approximation, but because
the charged particle source is not collimated, and annihilation rates at
our energies are not strongly angle-dependent, its use is justified.
We assume that the mean charged-particle velocity is zero and ignore their
spatial transport.  This is justified if they are trapped by a strong
magnetic field, as is likely near a neutron star (flare-like models of
gamma-ray bursts require high magnetic fields; see Katz 1982).  The
resulting equation for the charged-particle density is
$${\partial n_e \over \partial t} = \int I(E,\mu) I(E^\prime,\mu^\prime)
Q(E,E^\prime,\mu,\mu^\prime) dE dE^\prime d\mu d\mu^\prime - n_e^2 W(T),
\eqno(14)$$
where $Q$ is the azimuthally averaged pair-production coefficient and $W$ is
the thermally-averaged annihilation coefficient; we assume zero baryon
density so that the electron and positron densities are each $n_e$.  The
matter temperature $T$ is determined from the mean energy per charged
particle, which is found from the conservation of energy, using the mean
energy of the pairs produced and of the pairs annihilated.  Additional
details are presented by Carrigan (1987).

Because our purpose was to study nonlinear radiative transport as cleanly
as possible, we only considered pair production
and two- and three-photon pair annihilation.  Compton
scattering would destroy any spectral lines (for optical
depths greater than unity) and tend to thermalize the
emergent spectrum.  However, the model source spectra
considered lacked spectral lines, and the processes
considered also act to thermalize the spectra; therefore, Compton scattering
would not be likely to make a qualitative difference.
The inclusion of a spectral softening process (three-photon pair annihilation)
is important; in the absence of such a process, the mean particle
energy could not change.  There are many softening
processes; again, the precise one or ones chosen should make little
difference either at low optical depths (for which they are unimportant) or
at high optical depth (for which memory of the quantitative form of the
softening process is lost).  This justifies our neglect of the process
$\gamma + \gamma \rightarrow e^+ + e^- + \gamma$, which is complementary
(and comparable) to $e^+ e^- \rightarrow \gamma + \gamma + \gamma$ which we
do include.
\bigskip
\centerline{III. METHOD}
\bigskip
The nonlinear nature of the coupled radiative transfer equation (12) and the
electron density equation (14) poses problems for their solution.
Eleven spatial zones, logarithmically spaced in
radius out to ten times R, the inner radius, were
employed.  Numerical instability would occur in an explicit integration of
the transport terms if the time step employed were such that a photon
could traverse more than one spatial zone in one step ({\it i.e.}, for $\Delta
t \geq 0.259 R/c$ for the innermost zone); in order to improve the accuracy of
the time-dependent solution, we used a substantially smaller $\Delta t = 0.05
R/c$.  A fully explicit treatment of the problem is impractical,
except at low optical depths, because $\Delta t$ must also
be chosen small enough that the interactions do not
destroy all the photons (or all the charged particles) in
a single step.  At low optical depths, this condition is
less stringent than that imposed by the transport terms.
However, at high optical depths, the time steps permitted by an explicit
solution become very small, because the particle and photon densities are
large, and their interaction rates (per particle or photon) are proportional
to their densities, and the computational effort increases rapidly.

In practice, we used a semi-implicit technique, in which the transport terms
were evolved explicitly in time, while the local interaction terms
were solved implicitly, by matrix inversion.  This semi-implicit
method is not unconditionally stable, but it is
much more stable than explicit methods.  In addition, the
semi-implicit method required less memory than would be
needed for a fully implicit technique.  One factor which
has been found to influence stability is the temporal
dependence of the source term $S(E,r)$: if a strong source is
turned on suddenly, the semi-implicit method is made unstable
by the rapidly rising photon and particle densities.
We therefore turned the source on gradually, with $S$ proportional to $1 -
\exp(-t/t_0)^2$, with $t_0 = R/c$.  The desired steady state solutions are
obtained for $t \gg t_0$.

     The semi-implicit method consists of the following:
a transport equation
$${\partial \psi \over \partial t} = G[r,t,v,E,{\hat \Omega},\psi] - v{\hat
\Omega} \cdot \nabla \psi \eqno(15)$$
is approximated by the difference equation
$${\psi(t+\Delta t) - \psi(t) \over \Delta t} = G[r,t,v,E,{\hat
\Omega},\psi(t+\Delta t)] - v {\hat \Omega} \cdot \nabla \psi(t), \eqno(16)$$
with ${\hat \Omega} \cdot \nabla \psi$ approximated by the appropriate
(upstream) explicit differencing scheme.  The difference equation is
linearized in the quantity $\Delta \psi = \psi(t+\Delta t) - \psi(t)$ and then
solved for $\Delta \psi$:
$$\Delta \psi = \left[ 1 - {\partial G \over \partial \psi}\Delta t
\right]^{-1}\left(G - v {\hat \Omega} \cdot \nabla \psi \right) \Delta t.
\eqno(17)$$
The field $\psi$ will in general depend on $E$, $\hat \Omega$, and $r$, as
well as on time; as a result, the equation must be discretized
in these other quantities as well.  Write
$$\psi^i_j = \psi\left(r=r^i, E=E_j, {\hat \Omega}={\hat \Omega}_j, \ldots
\right), \eqno(18)$$
with the index $i$ standing for purely positional information
and $j$ and $k$ standing for everything else.  Then define
$$A^i_{jk} \equiv \delta_{jk} - {\partial G^i_j \over \partial \psi^i_k}.
\eqno(19)$$
The $A^i_{jk}$ may be regarded as forming the components of a set of matrices
$A^i$.  Then
$$\Delta \psi^i = \left( A^i \right)^{-1} \left(G^i - v{\hat \Omega} \cdot
\nabla \psi^i \right) \Delta t. \eqno(20)$$
In the present problem, $\psi$ refers to both the radiation
intensity $I$ and the charged particle density $n_e$.  The
index $i$ labels the radial zones ($r = r_i$), while
the indices $j$ and $k$ refer to energy,
direction, and species (photon or charged particle).  The
function $G$ includes an explicit source of photons as well
as terms corresponding to pair production by photons and
two- and three-photon pair annihilation.  The spatial and
angular derivative terms ${\hat \Omega} \cdot \psi$ were written solely in
terms of $\psi(t)$; were these terms also linearized in $\Delta \psi$,
one would arrive at a fully-implicit algorithm.  For each
radial zone $r_i$, the matrix $A^i$ is inverted, and $\Delta \psi^i$ is
found; this process is repeated at each radial zone, and
at each time-step, until convergence to a steady state is
achieved.

     The photon source spectrum was usually
taken to have a power-law form in energy, and
to fall off as $r^{-6}$.  Such a spatial dependence
might be expected if the burst were driven by magnetic reconnection, as is a
Solar flare, because a dipole field's energy density has this dependence,
or if the field's stress were to limit
the confinement of the source's energy density.  In general, models which
require the diffusion of near-equilibrium radiation through an opaque
atmosphere (a class which includes both thermonuclear and accretional impact
models) naturally produce black body spectra with $T \sim 10^{7\ \circ}$K,
inconsistent with the much harder, probably nonthermal (Katz 1985) spectra
of gamma-ray bursts.  Nonthermal models usually require that the magnetic
field produce the radiation or confine the radiating particles (Katz 1982),
justifying our use of an $r^{-6}$ source dependence.

     Eight energy and eight angular bins were employed.
The angular bins were chosen to be those of the eight-point
Gauss-Legendre quadrature scheme; the energy bins
were spaced logarithmically.
\bigskip
\centerline{IV. RESULTS}
\bigskip
Emergent steady-state spectra are shown in Figures 1--5; unless
otherwise stated the source spectrum is of the form
$S \propto E^{-\nu}r^{-6}$ and normalized by the parameter $\beta$ (Equation
11).  Results for low values of $\beta$ are given in
Figure 1 (for $\nu =1$) and in Figure 2 (for $\nu =2$).  At very
low source strengths (low $\beta$), the optical depth is small, and most
photons escape the source region without interacting; the
pair density is low, and the emergent spectrum is
proportional to the source spectrum.  At somewhat higher
source rates, an apparent annihilation feature (either a
peak or a broad ``shelf'' in the spectrum) is formed, and
the high-energy portion of the spectrum becomes increasingly
depleted.  Results for $\nu =2$ and $\beta >1$ are shown in
Figure 3.  The depletion of high-energy photons increases
as $\beta$ increases.  One striking feature of the calculations
is that the high-energy portion of the emergent spectrum is almost
independent of $\beta$ in both shape and magnitude; as $\beta$ increases
from 10 to 10$^4$, the flux at 700 keV increases by only a factor of
about 10, while the flux at 1 MeV and above is essentially
constant for $1 < \beta < 10^4$.

     How sensitive is this result to the assumptions
which underlie the calculations (specific interactions
included, choice of the explicit source of photons,
{\it etc.})?  In order to answer this question, calculations
where performed with somewhat different assumptions.  In
one calculation, the charged-particle temperature was, at
each time step, reduced to half of what it was calculated
to be; this simulates the effect of charged particles
radiating a significant fraction of their energy as low-energy
photons (by synchrotron emission, for example).
The result is shown in Figure 4, together with the result
of the corresponding ``standard'' case ($\nu =2$).  In another
calculation, the source spectrum was taken to be an
unnormalized Wien spectrum corresponding to a photon
temperature of 511 keV; for photon energies $< 3$ MeV, this
distribution is much harder than the power-law ($\nu =2$)
spectrum used in the other computations.  The results of
this calculation, and of the corresponding ``standard''
case, are shown in Figure 5.

     At high values of $\beta$ the high-energy portion is
insensitive to the details of the source spectrum.  This
is illustrated in Figure 5: the emergent spectrum for a
Wien-distribution source is essentially identical to that
for a power-law source at energies above 300 keV.

     Although the emergent spectrum is insensitive to the
energy dependence of the source spectrum, it is quite
sensitive to changes in the spatial dependence of the source.
Photons produced near the star's surface have become
well collimated by the time they have gone several star radii
(this is a consequence of the spherical geometry; far
from the source, all photons move nearly radially outward); any
high-energy photons introduced far above
the star's surface, and which are also travelling
roughly radially outward, are
below the kinematic threshold (Equation 1) to pair
production.  Therefore, the optical
depth in a zone far above the star's surface is reduced
below what it would have been in slab geometry, for which there is no such
geometrical collimation effect.  By
contrast, a source confined to a thin shell at the surface can only
produce photons in a region which is effectively slab-symmetric,
so its pair-production optical depth is
not reduced, and very few high energy photons emerge.

     A source even more extended than our $r^{-6}$
source might be expected to yield a still larger high-energy flux.  The
investigation of such a source would require that the transport
calculations be extended to larger distances (since the effective
source radius grows as the source becomes more
spatially extended).  The short ($\sim 0.1$ ms) time-scales over which
the intensities of some bursts have been observed to vary
imply source regions no larger than $\sim 30$ km, excluding more
extended source models, at least for these objects.  In addition, the
upper bound on high-energy flux at the source is inversely
proportional to the source radius (because the optical
depth scales roughly as $n_\gamma r_e^2 R$, or, equivalently,
because of the scale invariance of the transport equations).
The observed flux is
thus only linearly proportional to the source radius;
the upper bound on source distance rises only as the square root of
the source radius.
\bigskip
\centerline{V. CONCLUSIONS}
\bigskip
     A comparison between the observed (initial) spectrum
of the 5 March 1979 event and two model calculations is
shown in Figure 6.  For the purposes of comparison, the computed results
are scaled to a neutron star 10 km in radius and 1 Kpc from the Solar
System.  This distance is the largest for which the calculated emergent flux
at $E > 511$ KeV could even approximately fit the data.  The low-energy
portions of the spectra are in poor agreement with one another, but the
low-energy emergent spectrum depends strongly on the
source spectrum (and, in any case, it is the high-energy
spectrum which has yielded upper bounds on the distance
to the source of this burst).  Above about 300 keV (where
wildly dissimilar sources yield almost identical fluxes),
the calculated and observed spectra are in better
agreement.  This agreement suggests that the source of
the 5 March 1979 event is no farther than about 1 kpc
from the Earth.  A larger source region (either from a
source which fell off more slowly than $r^{-6}$ or from a neutron
star larger than 10 km in radius) could be somewhat
farther away; however, because the distance limit scales only as
the square root of the source radius, it cannot be
relaxed significantly (a factor of 2 or 3) under any set
of reasonable assumptions consistent with the observed $\leq 0.15$ ms
rise-time for that burst.  Increasing the energy release
rate (and hence, $\beta$) would not increase the high-energy
flux (as is shown in Figure 3); increasing the
hardness of the source spectrum will also not increase
the flux significantly (as shown in Figure 5) for any
``reasonable'' source spectrum.

     This distance limit is larger, by nearly an order of
magnitude, than that given by Epstein (1985).  In
part, this difference results from Epstein's
use of the condition $\tau < 1$ to limit the high-energy flux,
while we considered transport at arbitrary optical
depths.  The larger part of the discrepancy, however, is
due to the difference in our assumed sources: Epstein
considered a source limited to the surface of a neutron
star, while we employed a source which extended above the
surface.  As mentioned previously, the use of a source
limited to the star's surface (or to a thin shell around
the star) results in a significantly smaller high-energy
flux, in broad agreement with Epstein's work.

     Are there any possible loopholes, ways in which the distance
limit presented here could be evaded?  Of the four possibilities
listed in the Introduction:
\item{1.} Diffusion through optically thick regions is found to be
ineffective.
\item{2.} Softening of the spectrum lets the radiant
energy emerge (the required fluxes are far below the Planck function,
justifying our neglect of stimulated emission), but does not relax a bound
imposed by the {\it observed} high-energy flux.  It does, however, explain
the unusually soft spectrum observed for the very luminous March 5, 1979
event (assuming the LMC identification to be correct).
\item{3.} Our calculations show that the suggested ``window'' in the pair
production opacity is not effectively used by the flux; it does not
self-collimate.
\item{4.} The sudden release of energy in the form of gamma-rays and
relativistic particles, even if initially uncollimated, will, if unconfined,
lead to relativistic outward streaming motion and an outwardly collimated
radiation field.  The kinematic threshold to pair production (Equation 1)
would then suppress the pair production rate and the optical depth,
permitting the escape of multi-MeV photons.  Values of magnetic fields
characteristic of pulsars would interfere with radial outflow, but a
fireball of particles and photons produced near the surface could oscillate
between magnetic hemispheres, and during its upward motion produce
collimated outward flow.  Alternatively, particle
production may occur on the narrow funnel of open field lines near the
magnetic pole of a high field neutron star; this funnel may be widened if
sufficient energy is released to distort the magnetic field configuration
and ``blow out'' part of the magnetosphere.

     Another way to evade the distance limit is based on
the pulse pile-up effect suggested by Laros {\it et al.} (1985):
a gamma-ray scintillation detector will count two (or more)
photons as one photon (with energy equal to the total
energy of all the photons) if they are received
within a short time interval (of the order of a
microsecond).  If the energy of a burst were concentrated
in a series of short ``micro-bursts'', rather than being
more or less smoothly varying with time, it may be
possible to account for most (or all) the observed high-energy
flux in terms of micro-bursts of soft photons.
The plausibility of this explanation rests, however, on
the response times of detector scintillators and on
the minimum variability time-scales of GRBs.  From
light travel time arguments, the latter quantity cannot
be much less than about 30 microseconds if the burst
source is the whole surface of a neutron star, while even
for radiation from a small ($\sim 1$ km) region on a neutron
star, the limit is about 3 microseconds.  Pulse pile-up
may, but need not, be important if GRBs involve hot spots
on neutron-star surfaces; if the source region covers the
entire surface of a neutron star, however, pile-up is
probably unimportant.  In any case, theorists ought to be
reluctant to claim the observations have been misinterpreted (as
would be the case if pile-up were significant) without
strong evidence.

It is therefore not possible to resolve definitively the apparent conflict
between the statistically attractive identification of the March 5, 1979
gamma-ray burst with N49 in the LMC and the pair production bound on its MeV
gamma-ray flux.  Relativistic outflow and collimation may be the most
attractive hypothesis, for they would also resolve the problem (Imamura and
Epstein 1987) of the near absence of reprocessed X-rays from the neutron
star's surface.  Alternatively, the paucity of reprocessed X-rays may be
explained as a consequence of self-absorption, if gamma-ray burst distances
are $\sim 100$ Kpc, as may be implied by data from the BATSE on GRO (Katz
1992).  Pulse pileup or an accidental positional coincidence cannot
be excluded.

Observations of MeV gamma-rays from other bursters (Matz, {\it et al.} 1985)
raise similar issues, especially if their sources are very distant.  Because
gamma-ray bursts are diverse, problems and solutions found in one class
of them (such as soft gamma repeaters like the March 5, 1979 event) need not
be applicable to all.

Throughout this paper we have assumed that the charged particles are
confined by a strong magnetic field.  Consequently, annihilation gamma-rays
are produced with zero mean momentum (which we approximate as an isotropic
source).  This constrains their transport, and from it follow the limits on
the escaping hard photon flux and on the inferred distances to gamma-ray
burst sources.

The GRO observed (Meegan, {\it et al.} 1992) the sources of gamma-ray
bursts to be isotropically distributed on the sky; when combined with the
distribution of burst fluences this leads to the inference that the sources
are distributed either in a very extended Galactic halo or at cosmological
distances.  At the luminosities implied by halo distances a neutron star's
magnetic field may not be sufficient to confine the pair gas; at
cosmological luminosities expected neutron star magnetic stresses are
insignificant.  An unconfined pair gas should therefore be considered.

An optically thick mixture of gas and radiation may be considered as a
single fluid with a combined equation of state; it is not necessary to
consider radiation transport in order to describe its flow.  At
semi-relativistic temperature the sound speed $c_s$ approaches $c/3^{1/2}$.
If produced impulsively the opaque fluid undergoes nearly adiabatic
expansion, cooling on a time scale $\sim r/c_s$.  In the course of this
expansion the thermal energy, residing in an isotropic distribution of
photon and charged particle momenta, is converted to radially collimated
motion whose mean velocity is at least semi-relativistic.  The pairs, now
adiabatically cooled to $k_B T \ll m_e c^2$, annihilate.  The photons are
described by the same low temperature $T$ in the fluid frame and are below
the pair production threshold; in the laboratory frame they are collimated
in the outward direction and may have multi-MeV energies.

The emergent spectrum depends on details of the equilibration process.  In
the simplest model photons and pairs are initially in equilibrium at
temperature $T_0$ (at optical depths $< \alpha^{-1}$ this may not be the
case, and the photon chemical potential may be negative rather than zero).
Adiabatic expansion only scales down the photon energy and temperature in the
distribution function, but does not change the spectral shape; the Doppler
shift
to the observer's frame scales them up again.  The total photon number and
energy are conserved in adiabatic expansion of a photon gas, so the emergent
spectrum (in the observer's frame) would be identical to that of the initial
black body.  Annihilation of the pairs increases the photon temperature by a
factor between 1 (for $k_B T_0 \ll m_e c^2$) and $(11/4)^{1/3}$ (for $k_B
T_0 \gg m_e c^2$).  Power law spectral tails cannot be explained by these
processes, and require departures from equilibrium.  For example,
nonequilibrium survival of pairs in the rarefying and cooling outflow
(requiring low densities) may lead to a high energy annihilation spectrum
with an abrupt cutoff at $2 \gamma m_e c^2$, where the usual Lorentz factor
$\gamma$ is obtained from the bulk expansion speed.  The value of $\gamma$
depends on $T_0$; $\gamma \sim k_B T_0 / m_e c^2$ if $k_B T_0 \gg m_e c^2$
and $\gamma \gg 1$ if the pair density is very low ($k_B T_0 \ll m_e c^2$ in
equilibrium), while an equilibrium gas with $k_B T_0 \sim m_e c^2$ leads to
$\gamma \sim O(1)$.

The expanding pair-photon plasma is optically thick, and does not become
thin until its radius $r \sim (\sigma N_p)^{1/2}$, where $N_p$ is the total
number of photons and particles impulsively produced (for steady energy
release use $N_p \sim {\dot N}r/c$ to obtain $r \sim \sigma {\dot N}/c$; the
smaller of these two estimates of $r$ is applicable).  Substitution of
$\sigma_{es} \sim r_e^2$ leads to very large values of $r$ ($\sim 10^{13}$ cm
for Galactic halo bursts and $\sim 10^{16}$ cm for cosmological bursts).
These values are deceptive---once the fluid-frame temperature
(varying $\propto r^{-1}$ in the adiabatic expansion of a relativistic
fluid) drops to values $\ll m_e c^2$ the photons fall below the
pair-production threshold and the pairs rapidly annihilate and disappear.
Although the column density of photons far exceeds $\sigma_{es}^{-1}$ they
stream outward freely, and the effective photosphere forms at $r \sim r_0
(k_B T_0 / m_e c^2)$.  The observed pulse of radiation has a duration $\sim
r / (\gamma^2 c) \sim r_0 m_e c /(k_B T_0) \sim 10^{-4}$ sec because of its
angular collimation to a width $\theta \sim 1/\gamma$.  It is clear that
models of this kind cannot explain the observed durations and complex
time-histories of gamma-ray bursts unless there is a continuing release of
energy throughout the observed burst duration, in the range $10^{-1}$--$10^3$
seconds.  Such a continued release is plausible in ``solar-flare'' models of
bursts (Ruderman 1975, Katz 1982) but may be difficult to accomodate in
catastrophic events (neutron star birth, coalescence, or death) at
cosmological distances.

The inference of Galactic halo distances $\sim 100$ Kpc to gamma-ray bursts
(Meegan, {\it et al.} 1992; Katz 1992) answers one of the chief objections to
magnetic flare models.  They have suffered from the difficulty of tapping the
magnetostatic energy of a magnetosphere whose source currents are in the
neutron star's interior.  Magnetic reconnection, the likely power supply to
a flare, cannot occur in usual models of neutron star magnetospheres in
which current flows only on open field lines.  Flares can occur if current
loops from the interior erupt into the magnetosphere.  This occurs readily
in ordinary fluid stellar atmospheres.  In neutron stars it requires
magnetic stresses sufficient to break the solid crust, whose strength
$\mu_0$ is uncertain but may be $\sim 10^{25}$ dyne/cm$^2$ in the outer
crust and $\sim 10^{27}$ dyne/cm$^2$ in the inner crust (Ruderman 1972).  A
magnetic stress of this value implies a characteristic flare energy $\sim
\mu_0 \ell^3 \sim 10^{40}$--$10^{42}$ erg, where $\ell \sim 10^5$ cm is the
crust's thickness.  This crude estimate is consistent with halo distances
(but not with Galactic disc distances $\sim 100$ pc formerly considered
likely).  The narrow range of magnetic fields inferred (Murakami, {\it et
al.} 1988; Fenimore, {\it et al.} 1988) from spectral data is also explained
if the magnetic stress in a burst is determined by the crustal strengths
found in a homogeneous population of neutron stars.  The deficiency of weak
bursts observed by GRO implies that the energy release is (broadly)
concentrated in events around the knee of the $\log N / \log S$ curve, which
corresponds to $\sim 10^{43}$ erg at 100 Kpc distance.

We thank F. Perkins and M. A. Ruderman for discussions and NASA grants
NAGW-592 and NAGW-2028 for support.
\par
\vfil
\eject
\def\refformat{\hangindent=20pt \hangafter=1}
\centerline{References}
\bigskip
\parindent=0pt
Carrigan, B. J. 1987, Ph. D. Thesis, Washington U., St. Louis.
\medskip
Carrigan, B. J., and Katz, J. I. 1984, {\it Astron. Exp.} {\bf 1}, 89.
\medskip
Carrigan, B. J., and Katz, J. I. 1987, {\it Ap. J.} {\bf 323}, 557.
\medskip
Cavallo, G., and Rees, M. J. 1978, {\it M. N. R. A. S.} {\bf 183}, 359.
\medskip
Cline, T. L. 1980, {\it Comments Ap.} {\bf 9}, 13.
\medskip
Epstein, R. I. 1985, {\it Ap. J.} {\bf 297}, 555.
\medskip
\refformat
Fenimore, E. E., Laros, J. G., Klebesadel, R. W., Stockdale, R. E., and
Kane, S. R. 1982, in {\it Gamma Ray Transients and Related Astrophysical
Phenomena}, eds. R. E. Lingenfelter, H. S. Hudson, D. M. Worrall (New
York: AIP) p. 201.
\medskip
Fenimore, E. E., {\it et al.} 1988, {\it Ap. J. (Lett.)} {\bf 335}, L71.
\medskip
Guilbert, P. W., and Stepney, S. 1985, {\it M. N. R. A. S.} {\bf 212}, 523.
\medskip
Helfand, D. J., and Long, K. S. 1979, {\it Nature} {\bf 282}, 589.
\medskip
Imamura, J. N., and Epstein, R. I. 1987, {\it Ap. J.} {\bf 313}, 711.
\medskip
Katz, J. I. 1982, {\it Ap. J.} {\bf 260}, 371.
\medskip
Katz, J. I. 1985, {\it Ap. Lett.} {\bf 24}, 183.
\medskip
Katz, J. I. 1987, {\it High Energy Astrophysics} (Menlo Park: Addison-Wesley).
\medskip
Katz, J. I. 1992, {\it Ap. Sp. Sci.} in press.
\medskip
\refformat
Laros, J. G., Fenimore, E. E., Fikani, M. M., Klebesadel, R. W., van der
Klis, M., and Gottwald, M. 1985, {\it Nature} {\bf 318}, 448.
\medskip
Liang, E. P., and Petrosian, V., eds. 1986, {\it Gamma-Ray Bursts} (New
York: AIP).
\medskip
\refformat
Matteson, J. L. 1983, in {\it Positron-Electron Pairs in Astrophysics}, eds.
M. L. Burns, A. K. Harding, R. Ramaty (New York: AIP) p. 292.
\medskip
\refformat
Matz, S. M., Forrest, D. J., Vestrand, W. T., Chupp, E. L., Share, G. H.,
and Rieger, E. 1985, {\it Ap. J. (Lett.)} {\bf 288}, L37.
\medskip
Meegan, C. A., {\it et al.} 1992, {\it Nature} {\bf 355}, 143.
\medskip
Murakami, T., {\it et al.} 1988, {\it Nature} {\bf 335}, 234.
\medskip
\refformat
Nolan, P. L., Share, G. H., Chupp, E. L., Forrest, D. J., and Matz, S. M.
1984, {\it Nature} {\bf 311}, 360.
\medskip
Ruderman, M. A. 1972, {\it Ann. Rev. Astr. Ap.} {\bf 10}, 427.
\medskip
Ruderman, M. A. 1975, {\it Ann. N. Y. Acad. Sci.} {\bf 262}, 164.
\medskip
Schmidt, W. K. H. 1978, {\it Nature} {\bf 271}, 525.
\medskip
Svensson, R. 1983, {\it Ap. J.} {\bf 270}, 300.
\medskip
\refformat
Teegarden, B. J. 1984, in {\it High Energy Transients in Astrophysics}, ed.
S. E. Woosley (New York: AIP) p. 352.
\par
\vfil
\eject
\centerline{Figure Captions}
\bigskip
Figure 1: Emergent spectra (in arbitrary units) for $\nu = 1$.  Dashed-dotted
curve: $\beta = 0.01$.  Dashed curve: $\beta = 0.03$.  Dotted curve: $\beta =
0.1$.  Solid curve: $\beta = 0.3$.
\medskip
Figure 2: Emergent spectra for $\nu = 2$.  Solid curve: $\beta = 0.03$.
Dashed-dotted curve: $\beta = 0.1$.  Dashed curve: $\beta = 0.3$.  Dotted
curve: $\beta = 1$.
\medskip
Figure 3: Emergent spectra for $\nu = 2$.  Solid curve: $\beta = 10$.
Dashed-dotted curve: $\beta = 100$.  Dashed curve: $\beta = 10^3$.  Dotted
curve: $\beta = 10^4$.
\medskip
Figure 4: Emergent spectra with and without additional cooling of the
electron-positron gas.  $\beta = 10$, $\nu = 2$.  Dotted curve: no
additional cooling.  Dashed curve: electron-positron temperature reduced to
half its nominal value.
\medskip
Figure 5: Emergent spectra for $\beta = 10^4$.  Dotted curve: Source
$\propto E^{-2}$ ($\nu =2$).  Dashed curve: Source $\propto E^2\exp(-E)$
(Wien-like source spectrum).
\medskip
Figure 6: Comparison of calculated flux (photons cm$^{-2}$ s$^{-1}$
keV$^{-1}$) for $\nu = 2$, $\beta = 10^4$ (dotted line) and $\beta = 10^3$
(dashed line) to the spectrum of the initial 4 seconds of the March 5,
1979 burst.  The calculated fluxes assumed a 10 km radius neutron star 1 Kpc
from the solar system.
\par
\vfil
\eject
\bye
\end